\documentclass[prb,twocolumn,floatfix]{revtex4}

\usepackage{graphicx}
\usepackage{amsmath}
\usepackage{amssymb}
\usepackage[colorlinks=true,citecolor=blue,linkcolor=blue]{hyperref}
\usepackage{amsfonts}
\usepackage{color,xcolor}
\usepackage{epstopdf}
\usepackage{braket}
\usepackage{bm}
\usepackage{bbold}

\renewcommand{\vec}[1]{\ensuremath{\boldsymbol{#1}}}

\begin{document}

\title{Excitonic complexes in anisotropic atomically thin two-dimensional materials: black phosphorus and TiS$_3$}
\date{\today}
\author{M. Van der Donck}
\email{matthias.vanderdonck@uantwerpen.be}
\author{F. M. Peeters}
\email{francois.peeters@uantwerpen.be}
\affiliation{Department of Physics, University of Antwerp, Groenenborgerlaan 171, B-2020 Antwerp, Belgium}

\begin{abstract}
The effect of anisotropy in the energy spectrum on the binding energy and structural properties of excitons, trions, and biexcitons is investigated. To this end we employ the stochastic variational method with a correlated Gaussian basis. We present results for the binding energy of different excitonic complexes in black phosphorus (bP) and TiS$_3$ and compare them with recent results in the literature when available, for which we find good agreement. The binding energies of excitonic complexes in bP are larger than those in TiS$_3$. We calculate the different average interparticle distances in bP and TiS$_3$ and show that excitonic complexes in bP are strongly anisotropic whereas in TiS$_3$ they are almost isotropic, even though the constituent particles have an anisotropic energy spectrum. This is also confirmed by the correlation functions.
\end{abstract}

\maketitle

\section{Introduction}

One of the most important properties of atomically thin two-dimensional (2D) materials is the fact that the Coulomb interactions between different charge carriers are very strong. As a result, the binding energies of excitons, trions, and biexcitons in 2D materials with a band gap can be up to two orders of magnitude larger than in conventional semiconductors\cite{sc1,sc2,sc3,sc4,sc5}. 2D transition metal dichalcogenides (TMDs) form a well-known class of materials in which these strongly bound excitons\cite{exc1,exc2,exc3,exc4,exc5}, trions\cite{exc3,tri1,tri2}, and biexcitons\cite{bi} were recently observed.

Another type of 2D semiconductor is monolayer black phosphorus (bP), also called phosphorene\cite{bp1,bp2} which, as opposed to TMDs, exhibits a highly anisotropic band structure\cite{band1,band2}. It is also well-suited for technological applications such as field effect transistors\cite{fet} and photodetector devices\cite{detect1,detect2}. Transition metal trichalcogenides\cite{tmt} form another class of anisotropic 2D semiconductors\cite{bandtmt}. Recently, monolayer TiS$_3$, the prototypical representative of this class, has been synthesized and proposed as a candidate for application in transistors\cite{tis1,tis2}. This material exhibits a peculiar anisotropic band structure in which the conduction band is flatter in the $k_x$-direction whereas the valence band is flatter in the $k_y$-direction. Thus the anisotropy directions of electrons and hole bands are different from each other. This is in contrast to bP in which both the conduction and the valence band are flatter in the $k_y$-direction. Both these anisotropic 2D materials show interesting properties such as linear dichroism\cite{dichro1,dichro2,dichro3} and Faraday rotation\cite{faraday}.

In the present paper we investigate the binding energy and structural properties of excitons, trions, and biexcitons in bP and TiS$_3$. We employ the stochastic variational method (SVM) using a correlated Gaussian basis\cite{svm1,svm2}. This approach was successfully used to describe the binding energy of excitons, trions, and biexcitons in semiconductor quantum wells\cite{sc4} and more recently in 2D TMDs\cite{analytic,multi,tmds}.

Our paper is organized as follows. In Sec. \ref{sec:Model} we present an outline of our model and the stochastic variational method. The numerical results are discussed in Sec. \ref{sec:Results}. In Sec. \ref{sec:Summary and conclusion} we summarize the main conclusions.

\section{Model}
\label{sec:Model}

The low-energy Hamiltonian for an $N$-particle excitonic complex can be written as
\begin{equation}
\label{ham}
H = \sum_{i=1}^N\left(\frac{\hbar^2k_{i,x}^2}{2m_i^x}+\frac{\hbar^2k_{i,y}^2}{2m_i^y}\right)+\sum_{i<j}^N\text{sgn}(q_iq_j)V(|\vec{r}_i-\vec{r}_j|),
\end{equation}
with $q_i$ and $m_{i,x(y)}$ the charge and effective mass in the $x(y)$-direction of particle $i$. The interaction potential $V(r)$ is, due to non-local screening effects, given by\cite{screening1,screening2,screening3}
\begin{equation}
\label{interpot}
V(r) = \frac{e^2}{4\pi\kappa\varepsilon_0}\frac{\pi}{2r_0}\left[H_0\left(\frac{r}{r_0}\right)-Y_0\left(\frac{r}{r_0}\right)\right],
\end{equation}
where $Y_0$ and $H_0$ are the Bessel function of the second kind and the Struve function, respectively, with $\kappa=(\varepsilon_1+\varepsilon_2)/2$ where $\varepsilon_{1(2)}$ is the dielectric constant of the environment above (below) the material, and with $r_0=d\varepsilon/(2\kappa)$ the screening length where $d$ and $\varepsilon$ are, respectively, the thickness and the dielectric constant of the material. For $r_0=0$ this potential reduces to the bare Coulomb potential $V(r)=e^2/(4\pi\kappa\varepsilon_0r)$. Increasing the screening length leads to a decrease in the short-range interaction strength while the long-range interaction strength is unaffected. For very large screening lengths $r_0\rightarrow\infty$ the interaction potential becomes logarithmic, i.e. $V(r)=e^2/(4\pi\kappa\varepsilon_0r_0)\text{ln}(r_0/r)$. In the above Hamiltonian we have assumed that the electron and hole bands are parabolic, which is a good approximation for the low-energy spectrum of the considered materials.

The Schr\"odinger equation for the few-particle system can not be solved exactly for trions and biexcitons, although a direct numerical solution can be found for excitons. Therefore, in order to calculate the energies of the different excitonic complexes described by the above Hamiltonian, we employ the SVM in which the many-particle wave function $\Psi(\vec{r}_1,\ldots,\vec{r}_N)$ is expanded in a basis of size $K$\cite{svm1,svm2}:
\begin{equation}
\label{basisexp}
\Psi_{S,M_S}(\vec{r}_1,\ldots,\vec{r}_N) = \sum_{n=1}^Kc_n\varphi_{S,M_S}^{n}(\vec{r}_1,\ldots,\vec{r}_N),
\end{equation}
where the basis functions are taken as correlated Gaussians:
\begin{equation}
\label{corrgaus}
\begin{split}
&\varphi_{S,M_S}^{n}(\vec{r}_1,\ldots,\vec{r}_N) = \mathcal{A}\left(e^{-(\vec{x}^TA_n^x\vec{x}+\vec{y}^TA_n^y\vec{y})/2}\chi^n_{S,M_S}\right),
\end{split}
\end{equation}
where $\vec{x}$ and $\vec{y}$ are vectors containing, respectively, the $x$-components and $y$-components of the different particles. The matrix elements $\left(A_n^{x(y)}\right)_{ij}$ are the variational parameters and form a symmetric and positive definite matrix $A_n^{x(y)}$. $\chi^n_{S,M_S}$ is the total spin state of the excitonic complex corresponding to the total spin $S$ and $z$-component of the spin $M_S$, which are conserved quantities. This total spin state is obtained by adding step by step single-particle spin states. Therefore, multiple total spin states belonging to the same $S$ and $M_S$ value are possible, as these can be obtained by different intermediate spin states. For excitons and biexcitons we consider the $(S,M_S)=(0,0)$ singlet state and for trions we consider the $(S,M_S)=(1/2,1/2)$ doublet state. Finally, $\mathcal{A}$ is the antisymmetrization operator for the indistinguishable particles. The matrix elements of the different terms of the Hamiltonian between these basis functions can be calculated analytically\cite{analytic}.

To find the best energy value, we randomly generate matrices $A_n^{x(y)}$ and a spin function $\chi^n_{S,M_S}$ multiple times. The wave function with the set of parameters that gives the lowest energy is then retained as a basis function, and we now have a basis of dimension $K=1$. Subsequently, we again randomly generate a set of parameters and calculate the energy value in the $K=2$ basis consisting of our previously determined basis function and the new trial basis function. This is repeated multiple times and the trial function that gives the lowest energy value is then retained as the second basis function. Following this procedure, each addition of a new basis function assures a lower variational energy value and we keep increasing our basis size until we reach convergence of the energy value. Here, we found that a basis size of $K=50$ for excitons and $K=250$ for trions and biexcitons results in an energy convergence of 0.001 $\mu$eV, 0.1 $\mu$eV, and 1 $\mu$eV, respectively. This procedure is explained in more detail in Ref. [\onlinecite{svm1}].

The binding energies for excitons, trions, and biexcitons are, respectively, given by $E_b^{exc} = -E^{exc}$, $E_b^{tri} = E^{exc}-E^{tri}$, and $E_b^{bi} = 2E^{exc}-E^{bi}$, where $E^{exc}$, $E^{tri}$, and $E^{bi}$ are, respectively, the exciton, trion, and biexciton energy.

We will calculate the correlation function between two particles $i$ and $j$, defined as
\begin{equation}
\label{corr}
C_{ij}(\vec{r}) = \frac{\braket{\Psi|\delta(\vec{r}_i-\vec{r}_j-\vec{r})|\Psi}}{\braket{\Psi|\Psi}},
\end{equation}
which is the probability distribution of particles $i$ and $j$ being separated by a vector $\vec{r}$ and therefore satisfies
\begin{equation}
\label{norm}
\int C_{ij}(\vec{r})d\vec{r} = 1.
\end{equation}
The average distance between particles $i$ and $j$ is then obtained by
\begin{equation}
\label{dist}
\braket{r_{ij}} = \int_{-\infty}^{\infty}\int_{-\infty}^{\infty}\sqrt{x^2+y^2}C_{ij}(x,y)dxdy.
\end{equation}
Analogously we define the interparticle distance in the $x$ and $y$ direction as
\begin{equation}
\label{distxy}
\begin{split}
&\braket{x_{ij}} = \sqrt{\int_{-\infty}^{\infty}\int_{-\infty}^{\infty}x^2C_{ij}(x,y)dxdy}, \\
&\braket{y_{ij}} = \sqrt{\int_{-\infty}^{\infty}\int_{-\infty}^{\infty}y^2C_{ij}(x,y)dxdy}. \\
\end{split}
\end{equation}

\section{Results}
\label{sec:Results}
\begin{figure}
\centering
\includegraphics[width=8.5cm]{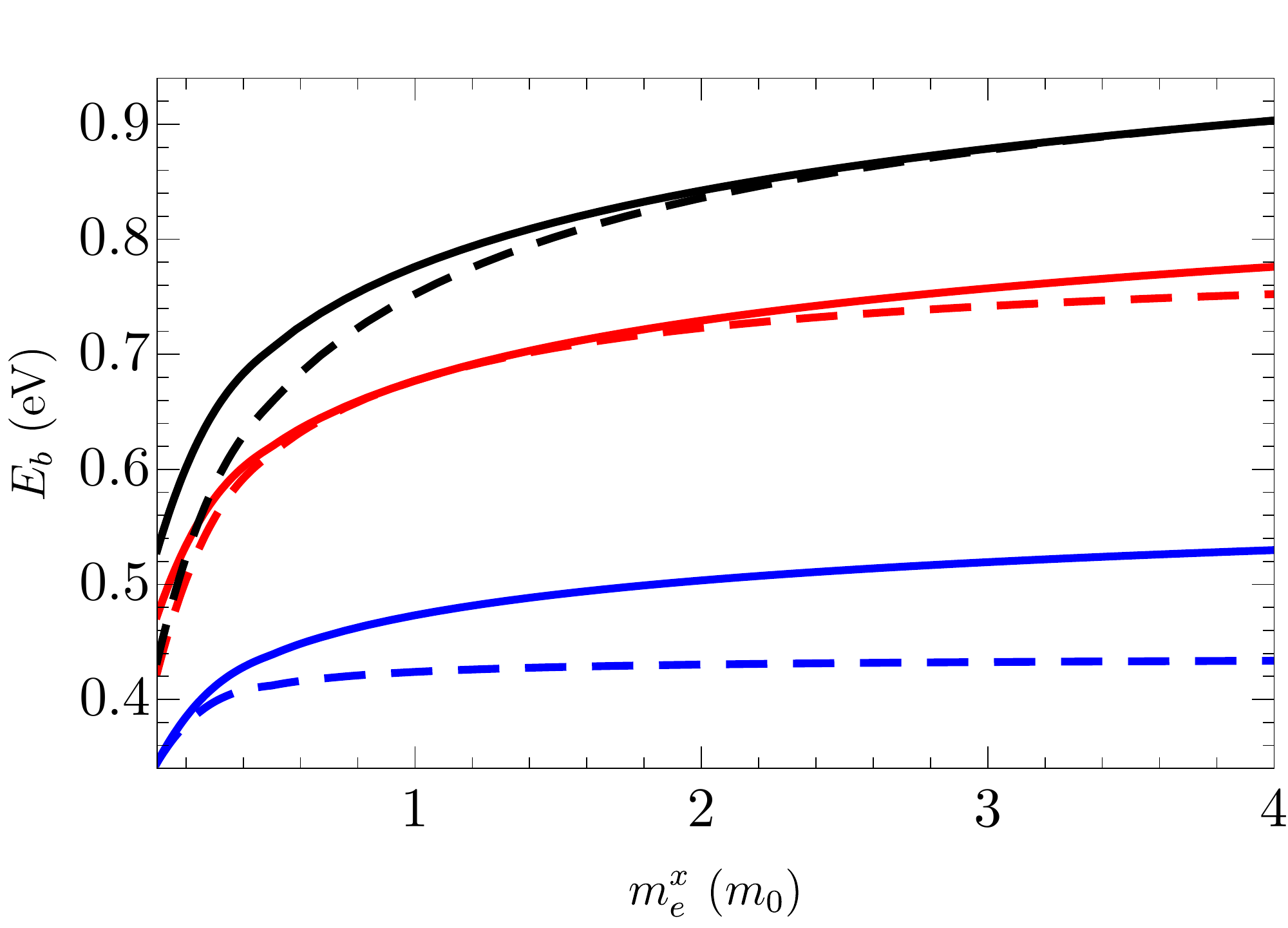}
\caption{(Color online) Exciton binding energy as a function of $m_e^x$ for $m_e^y=0.1m_0$ (blue), $m_e^y=m_0$ (red), and $m_e^y=4m_0$ (black) for $(m_h^x,m_h^y)=(m_e^x,m_e^y)$ (solid) and $(m_h^x,m_h^y)=(m_e^y,m_e^x)$ (dashed). We take $r_0=40$ \AA\ and $\varepsilon_b=\varepsilon_t=1$.}
\label{fig:massplot}
\end{figure}

In Fig. \ref{fig:massplot} we show the exciton binding energy for a general system as a function of the electron band mass in the $k_x$-direction for different values of the electron band mass in the $k_y$-direction. We study two distinct situations: $i$) identical electron and hole masses (solid curves) and $ii$) opposite electron and hole masses (the $k_x$-component of one equals the $k_y$-component of the other and vice versa) (dashed curves). We see that the binding energy for identical electron and hole masses is always larger than that for opposite masses except when the masses in the $k_x$- and $k_y$-direction are equal, i.e. at $m_e^x=0.1m_0$, $m_e^x=m_0$, and $m_e^x=4m_0$ for the blue, red, and black curves, respectively, as in this case the two situations are identical. This can be explained by the fact that the reduced mass $\mu^{x(y)}=m_e^{x(y)}m_h^{x(y)}/(m_e^{x(y)}+m_h^{x(y)})\leq\min(m_e^{x(y)},m_h^{x(y)})$, implying that in the opposite mass case $\mu^x=\mu^y$ will remain small when $m_e^x=m_h^y$ becomes large. Since excitonic properties are determined by the reduced mass and not the individual masses, this means that excitons in this system are isotropic and always 2D (even though the constituent particles are quasi-1D in the limit of large $m_e^x=m_h^y$) because of the limited reduced masses and therefore have limited binding energy. In the identical mass case, however, $\mu^{x(y)}=m_e^{x(y)}/2=m_h^{x(y)}/2$ and therefore $\mu^x$ increases linearly with increasing $m_e^x=m_h^x$ whereas $\mu^y$ remains constant. Excitons in this system are anisotropic and are quasi-1D in the limit of large $m_e^x=m_h^x$ and therefore their binding energy is large due to the additional confinement.
\begin{table}
\centering
\caption{Charge carrier masses\cite{mass} and screening lengths for bP and TiS$_3$.}
\begin{tabular}{c c c c c c c}
\hline
\hline
 & $m_e^x$ ($m_0$) & $m_e^y$ ($m_0$) & $m_h^x$ ($m_0$) & $m_h^y$ ($m_0$) & $r_0$ (\AA) \\
\hline
\hline
bP & 0.20 & 6.89 & 0.20 & 6.89 & 27.45 [\onlinecite{variational}] \\
\hline
TiS$_3$ & 1.52 & 0.40 & 0.30 & 0.99 & 44.34 [\onlinecite{tisscreen}] \\
\hline
\hline
\end{tabular}
\label{table:mattable}
\end{table}

For the remainder of the calculations we will use the parameters given in Table \ref{table:mattable}. In Table \ref{table:bindtable} we show the binding energies for excitons, negative trions, and biexcitons for bP and TiS$_3$ suspended in vacuum and on a SiO$_2$ substrate and compare them with other theoretical studies using diffusion Monte Carlo\cite{dmc}, the Numerov approach\cite{numerov}, a simple variational method\cite{variational}, and first-principles Bethe-Salpeter simulations\cite{tisscreen,bsedft,dftbse}. For bP, our results differ at most 16\% from those of Ref. [\onlinecite{dmc}]. More specifically for excitons the agreement is best with the results of Ref. [\onlinecite{bsedft}] (a difference of 0.3\% for vacuum) and least good with the results of Ref. [\onlinecite{variational}] (a difference of 21\% for SiO$_2$), which are obtained using a simple variational method. To the best of our knowledge, there are no other theoretical results for biexcitons in bP on a SiO$_2$ substrate. The binding energies for TiS$_3$ are in general smaller than those for bP. Our results agree very well, i.e. differing at most 9\%, with those from Ref. [\onlinecite{tisscreen}], which are obtained by numerically solving the relative Schr\"odinger equation either directly (excitons) or using an imaginary time evolution operator (trions). The authors also calculate the exciton binding energy for TiS$_3$ in vacuum by solving the Bethe-Salpeter equation and find 590 meV, which is in good agreement with our result. However, the result from Ref. [\onlinecite{dftbse}], which is obtained using first-principles Bethe-Salpeter simulations, differs almost a factor 2 from our result and those of Ref. [\onlinecite{tisscreen}].
\begin{table}
\centering
\caption{Exciton ($X$), negative ($X^-$) and positive ($X^+$) trion, and biexciton ($X_2$) binding energies (meV) for bP and TiS$_3$ for different substrates, compared with previous theoretical studies. We use $\varepsilon_r=3.8$ for SiO$_2$ and $\varepsilon_r=4.4$ for hBN (hBN$\times2$ denotes encapsulation in hBN).}
\begin{tabular}{cc|cc|cc}
\hline
\hline
 & Substrate & \multicolumn{2}{c}{bP} & \multicolumn{2}{|c}{TiS$_3$} \\
 &  & SVM & Theory & SVM & Theory \\
\hline
\hline
$X$ & Vacuum & 832.4 & 743.9 [\onlinecite{dmc}], 760 [\onlinecite{numerov}] & 537.1 & 560, 590 [\onlinecite{tisscreen}] \\
 & & & 710 [\onlinecite{variational}], 830 [\onlinecite{bsedft}] & & 920 [\onlinecite{dftbse}] \\
 & SiO$_2$ & 483.6 & 405.0 [\onlinecite{dmc}], 400 [\onlinecite{numerov}] & 314.7 & 330 [\onlinecite{tisscreen}] \\
 & & & 380 [\onlinecite{variational}] & & \\
 & hBN & 443.7 & - & 289.0 & - \\
 & hBN$\times2$ & 293.8 & - & 191.7 & - \\
\hline
$X^-$ & Vacuum & 56.3 & 51.6 [\onlinecite{dmc}] & 34.9 & 32 [\onlinecite{tisscreen}] \\
 & SiO$_2$ & 39.6 & 34.2 [\onlinecite{dmc}] & 25.3 & 23 [\onlinecite{tisscreen}] \\
 & hBN & 37.3 & - & 23.9 & - \\
 & hBN$\times2$ & 27.2 & - & 17.8 & - \\
 \hline
$X^+$ & Vacuum & 56.3 & 53 [\onlinecite{dmc}] & 34.0 & 36 [\onlinecite{tisscreen}] \\
 & SiO$_2$ & 39.6 & - & 24.2 & 26 [\onlinecite{tisscreen}] \\
 & hBN & 37.3 & - & 22.8 & - \\
 & hBN$\times2$ & 27.2 & - & 16.7 & - \\
 \hline
$X_2$ & Vacuum & 40.1 & 40.9 [\onlinecite{dmc}] & 25.8 & - \\
 & SiO$_2$ & 33.0 & - & 21.8 & - \\
 & hBN & 31.8 & - & 21.1 & - \\
 & hBN$\times2$ & 25.9 & - & 17.5 & - \\
\hline
\hline
\end{tabular}
\label{table:bindtable}
\end{table}

The results for both bP and TiS$_3$ on an hBN substrate are, due to the similar dielectric constant, close to those for a SiO$_2$ substrate. However, when the materials are encapsulated in hBN the binding energies of the excitonic complexes are considerably smaller. Furthermore, we find that the biexciton binding energy is almost always smaller than the trion binding energy. However, this difference becomes smaller with increasing substrate screening and eventually leads to the biexciton binding energy being larger than the positive trion binding energy for TiS$_3$ encapsulated in hBN. This is consistent with the general results of Ref. [\onlinecite{trionbiex}] which showed that both anisotropic band masses and a reduced screening length (in this case due to an increased $\kappa$) lead to an increase of the biexciton binding energy with respect to the trion binding energy. Finally, we find that the negative and positive trion binding energies are equal in bP because we assumed equal electron and hole band masses. In Ref. [\onlinecite{dmc}] a difference of 2.6\% between these two excitonic systems was found. For TiS$_3$ we find that the negative trion binding energy is larger than the positive trion binding energy, whereas the opposite behavior was found in Ref. [\onlinecite{tisscreen}].

There are only few experimental works studying excitonic systems in monolayer bP. An exciton binding energy of $900\pm120$ meV and $300$ meV was found in Ref. [\onlinecite{exp1}] and Ref. [\onlinecite{exp2}], respectively. Both studies used a SiO$_2$ substrate. The former differs about a factor 2 from our results (and even more from the other theoretical results), although it is remarkable that this value is in good agreement with the theoretical results for bP suspended in vacuum. It is possible that the experiment was accidentally performed on a part of the material which was lifted from the substrate. The result of Ref. [\onlinecite{exp2}] is in reasonable agreement with our result, i.e. a difference of 21\%. This study also found a trion binding energy of 100 meV, again on a SiO$_2$ substrate, which differs about a factor 3 from our result and that of Ref. [\onlinecite{dmc}]. To the best of our knowledge, there are no experimental results available for biexcitons in bP and for monolayer TiS$_3$ in general.
\begin{table}
\centering
\caption{Exciton, negative trion, and biexciton average interparticle distances (\AA), total and in the $x/y$-direction, for bP and TiS$_3$ suspended in vacuum.}
\begin{tabular}{c|ccc|ccc}
\hline
\hline
 & \multicolumn{3}{c}{bP} & \multicolumn{3}{|c}{TiS$_3$} \\
 & $r_{eh}$ & $r_{ee}$ & $r_{hh}$ & $r_{eh}$ & $r_{ee}$ & $r_{hh}$  \\
 & $x_{eh}$ & $x_{ee}$ & $x_{hh}$ & $x_{eh}$ & $x_{ee}$ & $x_{hh}$  \\
 & $y_{eh}$ & $y_{ee}$ & $y_{hh}$ & $y_{eh}$ & $y_{ee}$ & $y_{hh}$  \\
\hline
\hline
Exciton & 6.78 & - & - & 9.15 & - & -  \\
 & 8.17 & - & - & 7.76 & - & -  \\
 & 2.33 & - & - & 7.42 & - & -  \\
\hline
Trion & 12.18 & 20.06 & - & 15.22 & 24.32 & - \\
 & 16.07 & 23.37 & - & 13.11 & 18.30 & -  \\
 & 3.52 & 4.96 & - & 13.26 & 19.67 & -  \\
 \hline
Biexciton & 10.62 & 14.72 & 14.72 & 13.22 & 17.47 & 17.84 \\
 & 13.61 & 17.46 & 17.46 & 11.32 & 13.11 & 15.03  \\
 & 3.15 & 3.84 & 3.84 & 11.18 & 14.63 & 13.47 \\
\hline
\hline
\end{tabular}
\label{table:disttable}
\end{table}

In Table \ref{table:disttable} we show the average interparticle distances, total as well as resolved in the $x$/$y$-direction, for excitons, trions, and biexcitons in bP and TiS$_3$. In general, the interparticle distances are larger in TiS$_3$ as compared to bP, which is in correspondence with the smaller binding energies found in Table \ref{table:bindtable}. Excitons exhibit the smallest interparticle distance, as can be expected. More remarkably, trions show larger interparticle distances than biexcitons, even though their binding energy is larger. This is similar to what was found earlier in monolayer transition metal dichalcogenides\cite{analytic,tmds}. Furthermore, the average distance between particles of equal charge is larger than that between particles of opposite charge. Looking at the $x$/$y$-resolved interparticle distances, we see that the excitonic complexes in bP are strongly anisotropic, with the interparticle distances in the $x$-direction a factor 4-5 larger than those in the $y$-direction, whereas the excitonic complexes in TiS$_3$ are almost isotropic. It is also interesting to note that the electron-electron and hole-hole interparticles distances are identical in bP, due to the identical electron and hole band masses, whereas they are slightly different in TiS$_3$. More specifically, in TiS$_3$, the difference between the electron-electron and hole-hole interparticles distances in the $x$/$y$-direction is more pronounced than the difference between the total electron-electron and hole-hole interparticles distances. Furthermore, in the $x$-direction the electrons are located closer together than the holes whereas in the $y$-direction the opposite is true. This agrees with the band masses in Table \ref{table:mattable}, i.e. a larger electron band mass in the $x$-direction and a larger hole band mass in the $y$-direction.
\begin{figure}
\centering
\includegraphics[width=8.5cm]{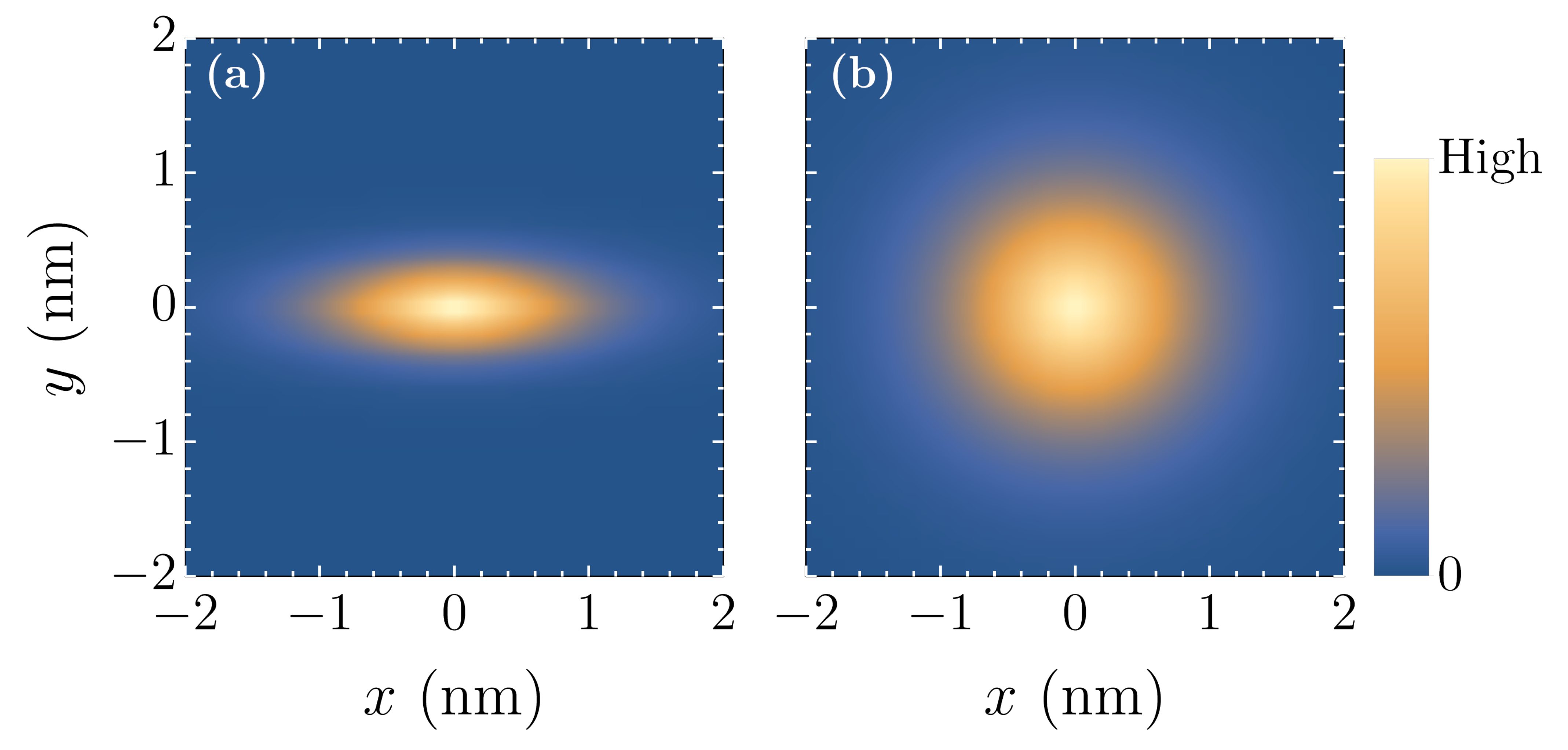}
\caption{(Color online) Electron-hole correlation functions for excitons in bP (a) and TiS$_3$ (b) suspended in vacuum.}
\label{fig:excplot}
\end{figure}
\begin{figure}
\centering
\includegraphics[width=8.5cm]{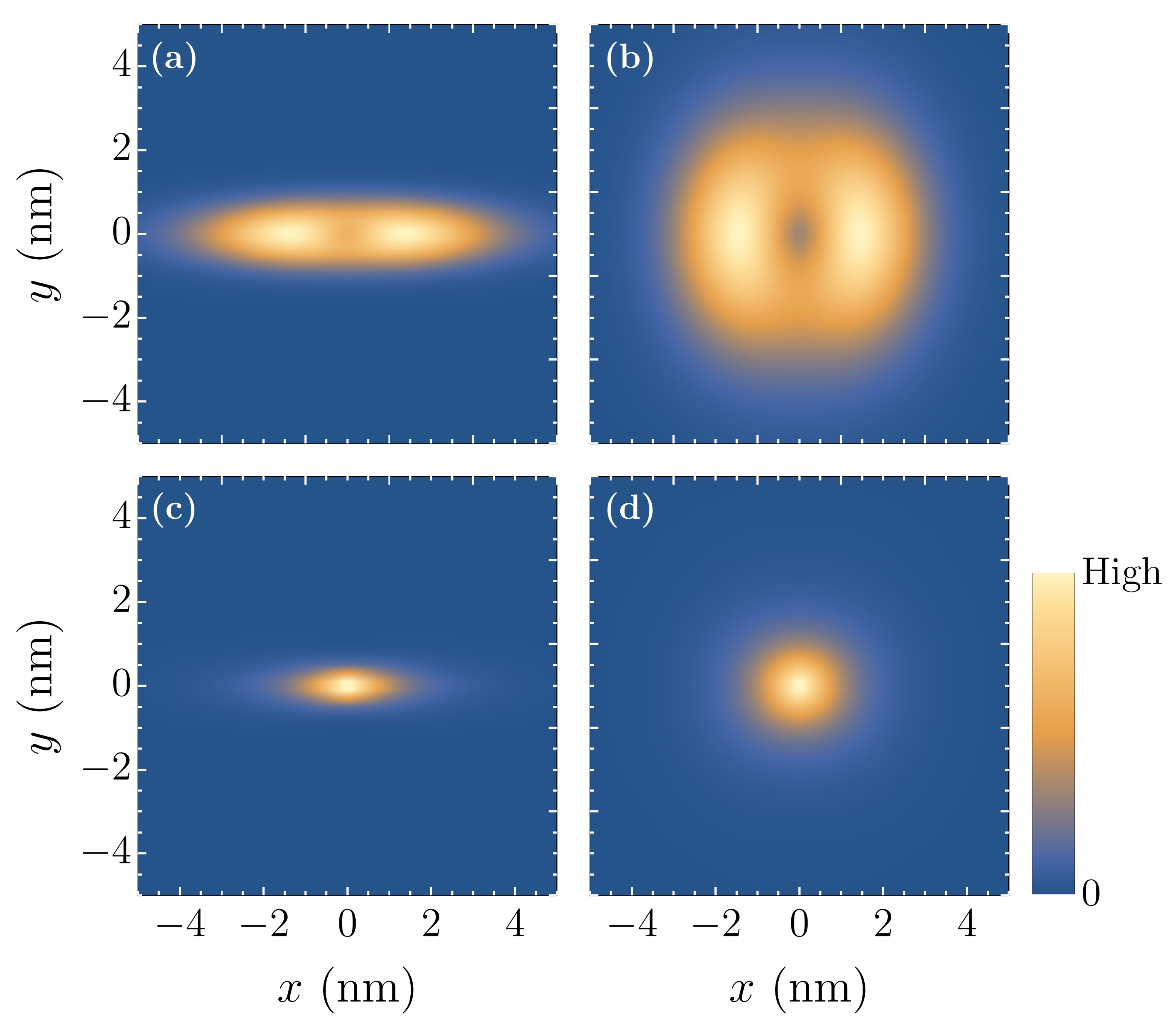}
\caption{(Color online) Electron-electron (a)-(b) and electron-hole (c)-(d) correlation functions for negative trions in bP (a)+(c) and TiS$_3$ (b)+(d) suspended in vacuum.}
\label{fig:triplot}
\end{figure}

Contour plots of the electron-hole correlation functions for excitons in bP and TiS$_3$ are shown in Fig. \ref{fig:excplot}. This clearly shows the strongly anisotropic behavior of excitons in bP and the almost isotropic excitons in TiS$_3$ as well as the fact that excitons in TiS$_3$ are in general larger than those in bP, even though in bP they are slightly more spread out in the $x$-direction.

We show the electron-electron and electron-hole correlation functions for negative trions in bP and TiS$_3$ in Fig. \ref{fig:triplot}. This again shows the difference in (an)isotropy between the two materials, although now the slight anisotropy in TiS$_3$ is also apparent in the electron-electron correlation function. The electron-electron correlation functions show two maxima along the $x$-direction, instead of one in the origin. This is a consequence of the Coulomb repulsion between the two electrons\cite{sc5,dmc} and this effect is therefore not present in the electron-hole correlation functions. This figure also clearly shows the larger spatial extent of the electron-electron correlation functions as compared to the electron-hole correlation functions, which is consistent with the average interparticle distances shown in Table \ref{table:disttable}.

\begin{figure}
\centering
\includegraphics[width=8.5cm]{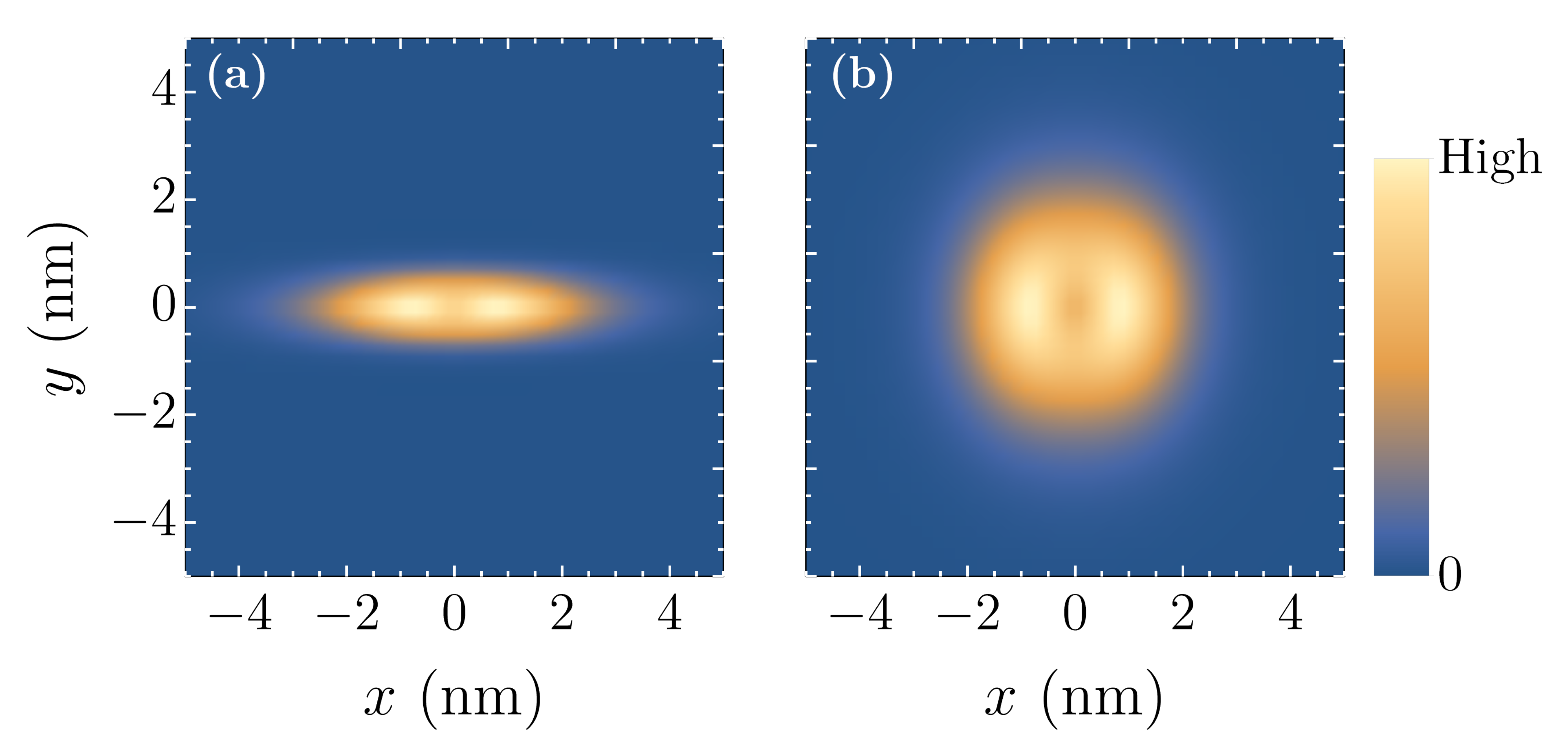}
\caption{(Color online) Electron-electron correlation functions for biexcitons in bP (a) and TiS$_3$ (b) suspended in vacuum.}
\label{fig:biplot}
\end{figure}

For biexcitons the electron-electron correlation functions for bP and TiS$_3$ are shown in Fig. \ref{fig:biplot}. This shows that the electron-electron correlation functions for biexcitons are very similar to those in negative trions, except that the system is more compact.

\section{Summary and conclusion}
\label{sec:Summary and conclusion}

In his paper, we studied the binding energy and structural properties of excitons, trions, and biexcitons in anisotropic 2D materials using the stochastic variational method with a correlated Gaussian basis and presented numerical results for bP and TiS$_3$.

We found that, in general, excitons in systems with equal electron and hole anisotropy have larger binding energies than those in systems with opposite electron and hole anisotropy, due to the anisotropy (isotropy) of the excitons in the former (latter) system. We also compared our results for the binding energy of different excitonic complexes in bP and TiS$_3$ with other theoretical works and found good agreement.

Furthermore, we calculated the different average interparticle distances and found that excitonic complexes in bP are strongly anisotropic, with the interparticle distances in the $x$-direction a factor 4-5 larger than those in the $y$-direction, whereas the excitonic complexes in TiS$_3$ are almost isotropic. We also found that the electron-electron and hole-hole interparticle distances for biexcitons in TiS$_3$ are slightly different due to the different band masses, which is most pronounced for the distances in the $x$/$y$-direction.

Finally, we calculated the correlation functions which clearly showed the anisotropic (isotropic) behavior of excitonic complexes in bP (TiS$_3$), as well as the effects of Coulomb repulsion between particles with equal charge.

\section{Acknowledgments}

This work was supported by the Research Foundation of Flanders (FWO-Vl) through an aspirant research grant for MVDD and by the FLAG-ERA project TRANS-2D-TMD.


\begin{thebibliography}{10}

\bibitem{sc1} R. J. Elliot, Phys. Rev. {\bf 108}, 1384 (1957).

\bibitem{sc2} V. D. Kulakovskii, V. G. Lysenk, and Vladislav B. Timofeev, Sov. Phys. Usp. {\bf 28}, 735 (1985).

\bibitem{sc3}
M. Hayne, C. L. Jones, R. Bogaerts, C. Riva, A. Usher, F. M. Peeters, F. Herlach, V. V. Moshchalkov, and M. Henini, Phys. Rev. B {\bf 59}, 2927 (1999).

\bibitem{sc4}
C. Riva, F. M. Peeters, and K. Varga, Phys. Rev. B {\bf 61}, 13873 (2000); {\it ibid.}, Phys. Status Solidi A {\bf 178}, 513 (2000).

\bibitem{sc5}
C. Riva, F. M. Peeters, and K. Varga, Phys. Rev. B {\bf 63}, 115302 (2001).

\bibitem{exc1}
T. Korn, S. Heydrich, M. Hirmer, J. Schmutzler, and C. Schller, Appl. Phys. Lett. {\bf 99}, 102109 (2011).

\bibitem{exc2}
G. Sallen, L. Bouet, X. Marie, G. Wang, C. R. Zhu, W. P. Han, Y. Lu, P. H. Tan, T. Amand, B. L. Liu, and B. Urbaszek, Phys. Rev. B {\bf 86}, 081301 (2012).

\bibitem{exc3}
K. F. Mak, K. He, C. Lee, G. H. Lee, J. Hone, T. F. Heinz, and J. Shan, Nat. Mater. {\bf 12}, 207 (2013).

\bibitem{exc4}
K. He, N. Kumar, L. Zhao, Z. Wang, K. F. Mak, H. Zhao, and J. Shan, Phys. Rev. Lett. {\bf 113}, 026803 (2014).

\bibitem{exc5}
A. Chernikov, T. C. Berkelbach, H. M. Hill, A. Rigosi, Y. Li, O. B. Aslan, D. R. Reichman, M. S. Hybertsen, and T. F. Heinz, Phys. Rev. Lett. {\bf 113}, 076802 (2014).

\bibitem{tri1}
C. H. Lui, A. J. Frenzel, D. V. Pilon, Y.-H. Lee, X. Ling, G. M. Akselrod, J. Kong, and N. Gedik, Phys. Rev. Lett. {\bf 113}, 166801 (2014). 

\bibitem{tri2}
A. Srivastava, M. Sidler, A. V. Allain, D. S. Lembke, A. Kis, and A. Imamo$\breve{\text{g}}$lu, Nat. Phys. \textbf{11}, 141 (2015).

\bibitem{bi}
E. J. Sie, A. J. Frenzel, Y.-H. Lee, J. Kong, and N. Gedik, Phys. Rev. B {\bf 92}, 125417 (2015).

\bibitem{bp1}
F. Xia, H. Wang, and Y. Jia, Nat. Commun. \textbf{5}, 4458 (2014).

\bibitem{bp2}
A. Castellanos-Gomez, J. Phys. Chem. Lett. \textbf{6}, 4280 (2015).

\bibitem{band1}
A. N. Rudenko and M. I. Katsnelson, Phys. Rev. B \textbf{89}, 201408(R) (2014).

\bibitem{band2}
D. \c{C}ak\i r, H. Sahin, and F. M. Peeters, Phys. Rev. B \textbf{90}, 205421 (2014).

\bibitem{fet}
Y. Du, H. Liu, Y. Deng, and P. D. Ye, ACS Nano \textbf{8}, 10035 (2014).

\bibitem{detect1}
N. Youngblood, C. Chen, S. L. Koester, and M. Li, Nat. Photon. \textbf{9}, 247 (2015).

\bibitem{detect2}
M. Engel, M. Steiner, and P. Avouris, Nano Lett. \textbf{14}, 6414 (2014).

\bibitem{tmt}
J. Dai, M. Li, and X. C. Zeng, WIREs Comput. Mol. Sci. \textbf{6}, 211 (2016).

\bibitem{bandtmt}
J. A. Silva-Guill\'en, E. Canadell, P. Ordej\'on, F. Guinea, and R. Rold\'an, 2D Mater. \textbf{4}, 025085 (2017).

\bibitem{tis1}
J. O. Island, M. Buscema, M. Barawi, J. M. Clamagirand, J. R. Ares, C. S\'anchez, I. J. Ferrer, G. A. Steele, H. S. J. van der Zant, and A. Castellanos-Gomez, Adv. Opt. Mater. \textbf{2}, 641 (2014).

\bibitem{tis2}
J. O. Island, M. Barawi, R. Biele, A. Almaz\'an, J. M. Clamagirand, J. R. Ares, C. S\'anchez, H. S. J. van der Zant, J. V. \'Alvarez, R. D'Agosta, I. J. Ferrer, and A. Castellanos-Gomez, Adv. Mater. \textbf{27}, 2595 (2015).

\bibitem{dichro1}
J. Qiao, X. Kong, Z.-X. Hu, F. Yang, and W. Ji, Nat. Commun. \textbf{5}, 4475 (2014).

\bibitem{dichro2}
J. A. Silva-Guill\'en, E. Canadell, F. Guinea, and R. Rold\'an, ACS Photonics \textbf{5}, 3231 (2018).

\bibitem{dichro3}
J. O. Island, R. Biele, M. Barawi, J. M. Clamagirand, J. R. Ares, C. S\'anchez, H. S. J. van der Zant, I. J. Ferrer, R. D'Agosta, and A. Castellanos-Gomez, Sci. Rep. \textbf{6}, 22214 (2016).

\bibitem{faraday}
K. Khaliji, A. Fallahi, and T. Low, International Conference on Infrared, Millimeter, and Terahertz waves, (Copenhagen, Denmark, 2018).

\bibitem{svm1}
Y. Suzuki and K. Varga, {\it Stochastic Variational Approach to Quantum Mechanical Few-body Problems}, (Springer-Verlag, Berlin, 1998).

\bibitem{svm2}
J. Mitroy, S. Bubin, W. Horiuchi, Y. Suzuki, L. Adamowicz, W. Cencek, K. Szalewicz, J. Komasa, D. Blume, and K. Varga, Rev. Mod. Phys. \textbf{85}, 693 (2013).

\bibitem{analytic}
D. W. Kidd, D. K. Zhang, and K. Varga, Phys. Rev. B \textbf{93}, 125423 (2016).

\bibitem{multi}
M. Van der Donck, M. Zarenia, and F. M. Peeters, Phys. Rev. B \textbf{96}, 035131 (2017).

\bibitem{tmds}
M. Van der Donck, M. Zarenia, and F. M. Peeters, Phys. Rev. B \textbf{97}, 195408 (2018).

\bibitem{screening1}
A. V. Chaplik and M. V. Entin, Zh. Eksp. Teor. Fiz. \textbf{61}, 2496 (1971).

\bibitem{screening2}
L. V. Keldysh, JETP Lett. \textbf{29}, 658 (1979).

\bibitem{screening3}
P. Cudazzo, I. V. Tokatly, and A. Rubio, Phys. Rev. B \textbf{84}, 085406 (2011).

\bibitem{mass}
Y. Aierken, D. \c{C}ak\i r, and F. M. Peeters, Phys. Chem. Chem. Phys. \textbf{18}, 14434 (2016).

\bibitem{variational}
A. Castellanos-Gomez, L. Vicarelli, E. Prada, J. O. Island, K. L. Narasimha-Acharya, S. I. Blanter, D. J. Groenendijk, M. Buscema, G. A. Steele, J. V. Alvarez, H. W. Zandbergen, J. J. Palacios, and H. S. J. van der Zant, 2D Mater. \textbf{1}, 025001 (2014).

\bibitem{tisscreen}
E. Torun, H. Sahin, A. Chaves, L. Wirtz, and F. M. Peeters, Phys. Rev. B \textbf{98}, 075419 (2018).

\bibitem{dmc}
A. Chaves, M. Z. Mayers, F. M. Peeters, and D. R. Reichman, Phys. Rev. B \textbf{93}, 115314 (2016).

\bibitem{numerov}
A. S. Rodin, A. Carvalho, and A. H. Castro Neto, Phys. Rev. B \textbf{90}, 075429 (2014).

\bibitem{bsedft}
V. Tran, R. Soklaski, Y. Liang, and L. Yang, Phys. Rev. B \textbf{89}, 235319 (2014).

\bibitem{dftbse}
Z. Jiang, Y. Li, S. Zhang, and W. Duan, Phys. Rev. B \textbf{98}, 081408(R) (2018).

\bibitem{trionbiex}
M. Szyniszewski, E. Mostaani, N. D. Drummond, and V. I. Fal'ko, Phys. Rev. B \textbf{95}, 081301(R) (2017).

\bibitem{exp1}
X. Wang, A. M. Jones, K. L. Seyler, V. Tran, Y. Jia, H. Zhao, H. Wang, L. Yang, X. Xu, and F. Xia, Nat. Nanotechnol. \textbf{10}, 517 (2015).

\bibitem{exp2}
J. Yang, R. Xu, J. Pei, Y. W. Myint, F. Wang, Z. Wang, S. Zhang, Z. Yu, and Y. Lu, Light: Sci. Appl. \textbf{4}, e312 (2015).

\end{thebibliography}
\end{document}